\documentclass[12pt]{amsart}

\usepackage{amsmath}
\usepackage{amsfonts}
\usepackage{amsthm}
\usepackage{amstext}

\theoremstyle{plain}
\newtheorem{thm}{Theorem}[section]
\newtheorem{prop}[thm]{Proposition}
\newtheorem{lemma}[thm]{Lemma}
\theoremstyle{definition}
\newtheorem{defn}[thm]{Definition}
\theoremstyle{remark}
\newtheorem{remark}[thm]{Remark}
\newtheorem*{statement}{Statement}
\newtheorem*{acknowledgement}{Acknowledgement}

\newcounter{nummer}
\newenvironment{araliste}{
\begin{list}
{\mbox{\textup{(\arabic{nummer})}}}
{\usecounter{nummer}\setlength{\leftmargin}{1cm}\setlength{\labelwidth}{1cm}}}
{\end{list}}

\newenvironment{latinliste}{
\begin{list}
{\mbox{\textup{(\roman{nummer})}}}
{\usecounter{nummer}\setlength{\leftmargin}{1cm}\setlength{\labelwidth}{1cm}}}
{\end{list}}

\newcommand{\supp}{\operatorname{supp}}
\newcommand{\dist}{{\operatorname{dist}}}
\newcommand{\N}{{\mathbb N}}
\newcommand{\R}{{\mathbb R}}

\numberwithin{equation}{section}

\begin{document}

\title{The electron density is smooth away from the nuclei}

\author[S. Fournais, M. and T. Hoffmann-Ostenhof and 
       T. \O. S\o rensen]{S\o ren Fournais, Maria Hoffmann-Ostenhof, 
       Thomas Hoffmann-Ostenhof and
       Thomas \O stergaard S\o rensen}

\address[S. Fournais, T. Hoffmann-Ostenhof, T. \O. S\o rensen]{The Erwin 
         Schr\"odinger International 
         Institute for Mathematical Physics                         \\
         Boltzmanngasse 9                                           \\
         A-1090 Vienna                                              \\
         Austria} 
\address[S. Fournais, present address]{Laboratoire de 
         Math\'{e}matiques                                          \\  
         Universit\'{e} Paris-Sud - B\^{a}t 425                     \\
         F-91405 Orsay Cedex                                        \\
         France}
\address[M. Hoffmann-Ostenhof]{Institut f\"ur Mathematik            \\
         Strudlhofgasse 4                                           \\
         Universit\"at Wien                                         \\
         A-1090 Vienna                                              \\
         Austria}
\address[T. Hoffmann-Ostenhof]{Institut f\"ur Theoretische Chemie   \\
         W\"ahringer\-strasse 17                                    \\
         Universit\"at Wien                                         \\
         A-1090 Vienna                                              \\
         Austria} 
\address[T. \O. S\o rensen, present address]{Department of 
         Mathematical Sciences                                      \\
         Aalborg University                                         \\
         Fredrik Bajers Vej 7G                                      \\
         DK-9220 Aalborg East                                       \\
         Denmark}
\email[S. Fournais]{fournais@imf.au.dk}
\email[M. Hoffmann-Ostenhof]{mhoffman@esi.ac.at}
\email[T. Hoffmann-Ostenhof]{thoffman@esi.ac.at}
\email[T. \O. S\o rensen]{sorensen@math.auc.dk}

\thanks{Work supported by the
        Carlsberg Foundation, European Union TMR grant FMRX-CT 96-0001,
        Ministerium f\"ur Wissenschaft und Verkehr der Republik 
        \"Osterreich, 
        and the Austrian Science Foundation, 
        grant number P12864-MAT}


\date{\today}

\begin{abstract}
  We prove that the electron densities of electronic eigenfunctions of atoms
  and molecules are smooth away from the nuclei.
\end{abstract} 

\maketitle

\section{Introduction and Statement of the Results.}

We consider an $N$-electron molecule with $L$ fixed nuclei whose 
non-relativistic Hamiltonian is given by 
\begin{multline}
  \label{Hmol}
  H_{N,L}(\mathbf R,\mathbf Z)=\sum_{j=1}^N\left({}-\Delta_j-\sum_{l=1}^L
  \frac{Z_l}{|x_j-R_l|}\right)\\
  +\sum_{1\le i<j\le N}\frac{1}{|x_i-x_j|}+ \sum_{1\le l<k\le L}\frac{Z_lZ_k}
  {|R_l-R_k|},
\end{multline}
where $\mathbf R=(R_1,R_2,\dots ,R_L)\in\mathbb R^{3L}$, \(R_{l}\neq R_{k}\)
for \(k\neq l\),
denote the positions of the $L$ nuclei
whose positive charges are given by $\mathbf Z=(Z_1, Z_2,\dots,Z_L)$.
The positions of the $N$ electrons are denoted by 
$(x_1,x_2,\dots,x_N)\in\mathbb R^{3N}$ where $x_j$ denotes the position
of the $j$-th electron in $\mathbb R^3$ and $\Delta=\sum_{j=1}^N\Delta_j$ is 
the $3N$-dimensional Laplacian.

The operator $H_{N,L}(\mathbf R,\mathbf Z)$ depends parametrically on 
$\mathbf R,\mathbf Z, L, N$ and
the last term in \eqref{Hmol} corresponds to the internuclear repulsion
which is just an additive term. It  will play no role in this paper and we 
will hence neglect it, i.e. from now on we let
$$
  H=H_{N,L}(\mathbf R, \mathbf Z)-{}\text{ internuclear repulsion}.
$$

The operator $H$ is selfadjoint on $L^2(\mathbb R^{3N})$ with operator domain 
$\mathcal D(H)=W^{2,2}(\mathbb R^{3N})$ and quadratic form domain 
$\mathcal Q(H)=W^{1,2}(\R^{3N})$, see e.g. Kato~\cite{Kato}.  

We consider the eigenvalue problem
\begin{equation}
  \label{Schroe}
  H\psi=(-\Delta +V)\psi=E\psi,
\end{equation}
where the potential $V$ is given by
\begin{equation}
  \label{potential}
  V=-\sum_{j=1}^N\sum_{l=1}^L\frac{Z_l}{|x_j-R_l|}+\sum_{1\le i<j\le N}\frac{1}
  {|x_i-x_j|}.
\end{equation}
Of course the eigenfunction $\psi$ and the associated eigenvalue  $E$ depend
parametrically on $\mathbf{R,Z},L,N$. 
Since the potential $V$ has singularities for $x_j=R_l$ and for  $x_i=x_j$
one cannot expect classical solutions. In 1957 Kato
showed \cite{Kato57} that any local solution $\psi$ to \eqref{Schroe} 
is locally Lipschitz, i.e.
\begin{equation*}
  \psi\in C^0_{\text{loc}}(\mathbb R^{3N})\text{ and } 
  |\nabla \psi|\in L^\infty_{\text{loc}}(\mathbb R^{3N}).
\end{equation*}
Kato also characterized the behaviour of such a solution near points
where two particles are close to each other (Cusp conditions).
(Extensions of these  results on the regularity of $\psi$ can be found in 
M. and T. Hoffmann-Ostenhof and Stremnitzer~\cite{Stremnitzer} and
M. and T. Hoffmann-Ostenhof and \O stergaard S\o rensen~\cite{firstWien}).
Of course, away from the singularities of $V$, any  local solution $\psi$
of \eqref{Schroe} is smooth by  elliptic regularity.

Equation \eqref{Schroe} is a partial differential equation in $3N$ variables
and hence only certain one-electron cases can be solved analytically.
(One-electron atoms and diatomic one-electron molecules with equal nuclear
charges). 

Most of the information about bound states of atoms and molecules that
Chemists and Physicists want to know is encoded in equation~\eqref{Schroe}.
(Of course one can go beyond \eqref{Schroe}, for instance allow for nuclear
motion,  include relativistic corrections, etc.). Since the electrons
are Fermions the physically relevant wavefunctions $\psi$ have to 
satisfy the Pauli principle. This amounts to requiring that $\psi$ 
transforms according to some specific irreducible representations of 
the symmetric group $\mathfrak S_N$. Our results will include this.

Already in the early times of quantum mechanics in the 1920's and 
the 1930's various attempts were made to replace the full 
$3N$-dimensional equation \eqref{Schroe} by simpler (usually non-linear) ones 
in 3 dimensions (Thomas-Fermi theory and Hartree-Fock theory). The 
relation of these approximations with the N-electron
Schr\"odinger equation has been analysed in detail for big atoms, see e.g. 
Lieb and Simon~\cite{LiSi77},
Lieb~\cite{Li81b}, and Lieb and Simon~\cite{LiSi77b}.

One important observation and motivation for the development of these
and other approximation schemes was the insight that in order to
calculate the energy $E$ or one- and two-electron operator expectation values
there is no  need  for the full wave function $\psi(x_1,x_2,\dots,x_N)$
but only for the {\bf one-electron density} 
${\mathbf \rho(x)}, x\in \mathbb R^3$, the
{\bf two-electron density} $\rho_2(x,x'),\ (x,x')\in\mathbb R^6$ and for the 
{\bf one-electron density matrix} $\gamma_{1}(x,x'),\ (x,x')\in\mathbb R^6$. 
These quantities are defined as follows: Defining 
\begin{align*}
  \psi_j(x)&=\psi(x_1,\dots,x_{j-1},x,x_{j+1},\dots,x_N)
  \intertext{ and } 
  \psi_{i,j}(x,x')&=
  \psi(x_1,\dots,x_{i-1},x,x_{i+1},\dots,x_{j-1},x',x_{j+1},\dots,x_N),
\end{align*}
and assuming without loss of generality that $\psi$ is real valued (since the 
coefficients of $H$ are real), the functions 
$\rho,\rho_{2}$, and $\gamma_{1}$ are given by
\begin{align}
  \label{rho}
  \rho(x)&=\sum_{j=1}^{N}\int \psi_j^2\,d\hat x_j,\\
  \label{rho2}
  \rho_2(x,x')&=\sum_{1\le i\neq j\le N}\int\psi_{i,j}^2\,d\hat x_{i,j},
  \intertext{and}
  \gamma_{1}(x,x')&=\sum_{j=1}^{N}\int\psi_j(x)\psi_j(x')\,d\hat x_j
  \label{gamma}.
\end{align}
Here  $d\hat x_j$, respectively $d\hat x_{i,j}$, means integration over 
all variables except $x_j$, repectively $x_i,x_j$. Of course 
$\rho(x)=\gamma_{1}(x,x)$. 

More recently very successful approximation schemes have been developped
called Density Functional Theories (DFT).
These schemes 
use  some non-linear functionals in which only the one-electron density 
\(\rho\) occurs and  lead to surprisingly good approximations to ground state
energies and molecular geometries (see
e.g. Eschrig~\cite{Eschrig}). However, their relation to the full  
Schr\"odinger equation remains unclear.

It is therefore  surprising that  the eigenfunction $\psi$ and in
particular the electron density $\rho$ itself,
defined in \eqref{rho}, have only rarely been the subject of  mathematical
analysis (see \cite{firstWien} and references therein).

Here we prove, based on recent work \cite{firstWien}, natural smoothness 
results about 
the quantities defined in \eqref{rho}, \eqref{rho2}, and \eqref{gamma}.

We do not assume anything about the eigenvalue $E$---in particular, it
could be an embedded eigenvalue. The only
assumption is that $\psi$ satisfies the following decay estimate: 
\begin{equation}
  \label{eq:decay}
  |\psi({\mathbf x})| \leq c e^{-\lambda |{\mathbf x}|} \text{ for all
    } {\mathbf x}\in\R^{3N},
\end{equation}
for some $c,\lambda > 0$.
\begin{remark}
  Since $\psi$ is continuous, (\ref{eq:decay})
  is only an assumption on the behaviour near infinity. For references
  on the exponential decay of eigenfunctions, see e.g. Simon~\cite{Si-semi}.
\end{remark}
\begin{remark}
  We only assume the exponential decay for simplicity. Our proofs
  extend to the case where the eigenfunction \(\psi\) decays faster than
  polynomially.   
\end{remark}

The main result of this paper is the following theorem:
\begin{thm}
\label{thm:density_smooth}
Let \(\psi\) be an eigenfunction of \(H\),
satisfying (\ref{eq:decay}). Let \(\rho, \rho_{2}\),
and \(\gamma_{1}\) be as defined in \eqref{rho}, \eqref{rho2}, and
\eqref{gamma}, and define furthermore \(n(x)=\rho_{2}(x,x)\).

Let \(D=\{(x,x)\in\R^{6}\}\subset\R^{6}\), and define
\begin{align*}
  \Sigma=\big(\{R_{1},\ldots,R_{L}\}\times\R^{3}\big)\cup 
  \big(\R^{3}\times\{R_{1},\ldots,R_{L}\}\big)\subset\R^{6}.
\end{align*}
Then
\begin{align}
 \rho, n&\in C^{\infty}(\R^{3}\setminus \{R_{1},\ldots,R_{L}\})
 \intertext{and}
 \rho_{2} &\in C^{\infty}(\R^{6}\setminus (\Sigma \cup D )), 
  \gamma_{1}\in C^{\infty}(\R^{6}\setminus\Sigma).
\end{align}
Furthermore, all the derivatives of $\rho$ satisfy an exponential
decay estimate near infinity: 
Let $\gamma\in\N^{3}$ be a multi-index. 
Then for all $\epsilon > 0$ there exists a constant $c = c(\gamma,
\epsilon)$ such that  
\begin{equation}
  \label{eq:rho_exponential}
  |\partial_{x}^{\gamma} \rho(x)| \leq c e^{-(\lambda-\epsilon) |x|}\quad
  \text{for all}\quad|x| > \max \{ |R_1|,\ldots, |R_L| \} + 1.
\end{equation}
\end{thm}

Similar results hold for $n$-electron densities 
and for $n$-electron density matrices.  

\begin{remark}
\label{rem:simplify}
For simplicity we will only prove Theorem~\ref{thm:density_smooth} for
atoms (i.e. for $L=1$, $R_1=0$) and only indicate the necessary
modifications for the molecular case. Moreover, we only explicitly
treat the density $\rho$, the proofs in the cases of $n, \rho_2$, and
$\gamma_1$ being essentially the same. Finally, it is enough to prove
smoothness of each term $\int \psi_j^2\,d\hat x_j$ in the sum
\eqref{rho}, and we therefore restrict ourselves to 
$$
\int \psi_1^2\,d\hat x_1 = \int \psi^2(x,x_2,\ldots,x_N)\,dx_2 \cdots dx_N.
$$
\end{remark}

\section{Additional regularity of eigenfunctions.}

We will need to know something about the regularity properties of
$\psi$ in order to conclude that $\rho$ is smooth.  
In order to study the regularity of $\psi$ it is convenient to work in
spaces of functions which are H\"older continuous. Let us recall the
definition of H\"older continuity: 
\begin{defn}
  Let \(\Omega\) be a domain in \(\R^{n}\), \(k\in\N\), and
  \(\alpha\in(0,1]\). We say that a function \(u\) belongs to
  \(C^{k,\alpha}(\Omega)\) whenever
  \(u\in C^{k}(\Omega)\), and for all \(\beta\in\N^{n}\) with
  \(|\beta|=k\), and all open balls $B(x_{0},r)$ with
  $\overline{B(x_{0},r)}\subset\Omega $, we have 
  \begin{align*}
    \sup_{x,y\in B(x_{0},r),\,x\neq y}
    \!\!\!\!\!\!\!\!\!
    \frac{|\partial^{\beta}u(x)-\partial^{\beta}u(y)|}{|x-y|^{\alpha}}
    \leq C(x_{0},r).
  \end{align*}
  For any domain \(\Omega'\), with \(\overline{\Omega'}\subset\Omega\),
  we now define the following norms: 
  \begin{align}
    \nonumber
    |u|_{C^{k,\alpha}(\Omega')}=\sum_{|\beta| \leq k}\|\partial^{\beta}
    u\|_{L^{\infty}(\Omega')} 
    +\sum_{|\beta| = k}\sup_{x,y\in\Omega',\,x\neq y}
    \frac{|\partial^{\beta}u(x)-\partial^{\beta} u(y)|}{|x-y|^{\alpha}}.
  \end{align}
\end{defn}

We will need the following result on elliptic
regularity in order to conclude
that the solutions of elliptic second order equations with
bounded coefficients are $C^{1,\alpha}$. The proposition is a
reformulation of Corollary 8.12 in Gilbarg and Trudinger~\cite{GandT},
adapted for our purposes:  
\begin{prop}
  \label{prop:GT}
  Let \(\Omega\) be a bounded domain in \({\mathbb
  R}^{n}\) and suppose \(u\in
  W^{1,2}(\Omega)\) is a weak solution of \(\Delta u
  +\sum_{j=1}^{n}b_{j}D_{j}u+Wu=g\) in \(\Omega\), where
  \(b_{j},W,g\in L^{\infty}(\Omega)\). Then \(u\in
  C^{1,\alpha}(\Omega)\) for all \(\alpha\in(0,1)\) and
  for any domain \(\Omega'\), \(\overline{\Omega'}\subset\Omega\) we have
  \begin{align*}
    |u|_{C^{1,\alpha}(\Omega')}\leq
    C\big(\sup_{\Omega}|u|+\sup_{\Omega}|g|\big)
  \end{align*}
  for \(C=C(n,M,\dist(\Omega',\partial\Omega))\), with
  \begin{align}
    \max_{j=1,\ldots,n}\{1,\|b_{j}\|_{L^{\infty}(\Omega)},
    \|W\|_{L^{\infty}(\Omega)},\|g\|_{L^{\infty}(\Omega)}\} \leq M.
    \nonumber
  \end{align}
\end{prop}

Our regularity result on $\psi$ is the following:
\begin{lemma}
\label{lem:smoothness}
Let $\psi$ be an $N$-electron atomic eigenfunction
satisfying the decay estimate (\ref{eq:decay}). 
Let $P$ and $Q$ be a partition of $\{1,\ldots,N\}$:
\begin{align*}
  \{1,\ldots,N\} = P \cup Q,\,\,\,\,\,\,\, P \cap Q = \emptyset,\,\,\,\,\,\,\,
  P \neq \emptyset.
\end{align*}
Define $x_P$ as 
\begin{align*}
  x_{P} = \frac{1}{\sqrt{|P|}}\,\sum_{j \in P} x_j \in {\mathbb R}^3,
\end{align*}
and let $T$ be any orthogonal transformation such that
$T(x_1,\ldots,x_N) = (x_P,x')$ with $x'\in {\mathbb R}^{3N-3}$. 

Let $\epsilon > 0$ be given and define $U_P \subset {\mathbb R}^{3N}$
as the open set: 
\begin{align}
  \label{eq:def_U_P}
  U_P = \Big\{ (x_1,\ldots,x_N) \in {\mathbb R}^{3N}\,\Big|&\, |x_j| > 
  \epsilon \mbox{ for } j \in P, \nonumber \\
  & |x_j-x_k| > \epsilon \mbox{ for } 
  j \in P, k \in Q \Big\}.
\end{align}
Then
\begin{align*}
  \partial_{x_P}^{\gamma} (\psi \circ T^*)  \in C^{0,1}(T
  U_P)\quad\text{ for all } 
  \gamma \in {\mathbb N}^{3}.
\end{align*}
Furthermore, the following a priori estimate holds:
\begin{align*}
  \big| \nabla \partial_{x_P}^{\gamma}
  (\psi\circ T^*)(x_P,x') \big| +
  \big| \partial_{x_P}^{\gamma}
  &(\psi\circ T^*)(x_P,x') \big| \nonumber \\
  &\leq C e^{-\lambda |(x_P,x')|} \,\,\,\,\,\,\,\,
  \text{ for all } (x_P,x') \in T U_P
\end{align*}
for some $C=C(\gamma)>0$.
\end{lemma}

\begin{remark}
  One could interpret $U_P$ as a (very large) neighbourhood of a
  singularity ${\mathbf x}^0$ of $V$ where the electrons with
  coordinates $x_j$ with $j \in Q$ sit on the 
  nucleus and the electrons with coordinates $x_j$ with $j \in P$ sit
  on each other 
  away from the nucleus, i.e. ${\mathbf x}^0 =
  (x_1^0,\cdots,x_N^0)$, with 
  \begin{eqnarray*}
    x_j^0 = 0 && \text{ for all } j \in Q, \\
    x_j^0 = y^0 \neq 0 &&  \text{ for all } j \in P.
  \end{eqnarray*}
  Notice that $x_P$ is (up to a scalar multiple) the centre of mass of
  the electrons $j$ with $j \in P$. Loosely speaking  
  Lemma~\ref{lem:smoothness} then says that in the neighbourhood $U_P$,
  $\psi$ is smooth with respect to the centre of mass coordinate $x_P$.

  One can also consider ${\mathbf x}^0$ as a two-cluster singularity---one
  group of electrons on each other at the nucleus, another group of
  electrons on top of each other away from the nucleus. It is, of
  course, possible to have many clusters. Lemma~\ref{lem:smoothness} is
  a special case of a more general theorem: If we have any number of
  clusters which are separated from each other and from the nucleus,
  then the eigenfunction $\psi$ can be differentiated any number of
  times with respect to the centre of mass of each cluster. 

  We only need the two-cluster version
  (see Lemma~\ref{lem:smoothness}) in the present paper. The more general
  result will be used in a detailed investigation of the regularity of
  the wavefunction near all kinds of singularities of the potential
  (see~\cite{second_regularity}). 
\end{remark}

\begin{remark}
  In the case of molecules the definition of $U_P$ is slightly different: 
  \begin{align*}
    U_{\text{MOL}} = \Big\{ (x_1,\ldots,x_N) \in {\mathbb R}^{3N}\,\Big|\,
    &\min_{1 \leq l \leq L} |x_j- R_l| >  
    \epsilon \mbox{ for } j \in P, \\
    & |x_j-x_k| > \epsilon \mbox{ for } 
    j \in P, k \in Q \Big\}.
  \end{align*}
  Apart from that, Lemma~\ref{lem:smoothness} remains unchanged.
\end{remark}
Before we prove Lemma~\ref{lem:smoothness}, let us fix
some notation. We may assume without loss of generality that $P =
\{1,\ldots,N_1\}$, with $N_1 \leq N$. Then any orthogonal
transformation $T$ satisfying the assumptions in the statement of
Lemma~\ref{lem:smoothness} can be written as: 
$$
  T = \left(\begin{array}{c}
  \begin{array}{cccccc}
    \frac{1}{\sqrt{N_1}}& \cdots& \frac{1}{\sqrt{N_1}}& 0 & \cdots & 0
    \end{array} \\
    \, \\ \, \\
    \tilde{T} \\ \, \\ \,
  \end{array}
  \right),
$$
with the first row being understood as $3\times 3$ matrices---first
$N_1$ repetitions of $\frac{1}{\sqrt{N_1}}I_3$ and then $N-N_1$
repetitions of the $3 \times 3$ $0$-matrix. The remaining part of the
matrix, $\tilde{T} \in 
M_{3N-3,3N}({\mathbb R})$ is such that the complete matrix $T$ is
orthogonal. We will denote the $(3N-3)\times3$-columns of $\tilde{T}$
by $t_j$, i.e. 
$$
  \tilde{T} = \left(
  \begin{array}{ccc}
    t_1&\cdots&t_N \\
  \end{array} \right) ,
$$
with $t_j \in M_{3N-3,3}({\mathbb R})$. Then we get:
\begin{equation}
  \label{eq:T_stjerne}
  T^* = \left( 
  \begin{array}{cccccc}
    \frac{1}{\sqrt{N_1}}&&& t_1^*& & \\
    \vdots && &\vdots &&\\
    \frac{1}{\sqrt{N_1}}&&& t_{N_1}^*& \\
    0 &&& t_{N_1+1}^*& \\
    \vdots &&& \vdots& \\
    0 &&& t_{N}^* &
  \end{array} \right).
\end{equation}

\begin{proof}
For the proof of Lemma~\ref{lem:smoothness} we first proceed as 
in M. and T. Hoffmann-Ostenhof and \O stergaard S\o rensen~\cite{firstWien}:
We make the `Ansatz'
\begin{align}
  \label{eq:ansatz}
  \psi = e^{F-F_1}\psi_1,
\end{align}
with
\begin{align*}
  F= \sum_{j=1}^N{}-\frac{Z}{2} |x_j| + \sum_{1\leq j< k\leq N} 
  \frac{1}{4}|x_j-x_k|,
\end{align*}
and
\begin{align*}
  F_1 = \sum_{j=1}^N{}-\frac{Z}{2} \sqrt{|x_j|^2+1} + \sum_{1\leq j< 
  k\leq N} \frac{1}{4}\sqrt{|x_j-x_k|^2+1}.
\end{align*}
Observing (see \eqref{potential} with $L=1$, $R_1 = 0$), 
\begin{equation}
  \label{eq:choice_of_F}
  \Delta F = V,
\end{equation}
we get the following equation for $\psi_1$ using \eqref{Schroe} and
(\ref{eq:ansatz}): 
\begin{equation}
  \label{eq:eq_for_psi1}
  \Delta \psi_1 + 2\nabla(F-F_1)\cdot \nabla \psi_1 
  +\big(|\nabla(F-F_1)|^2-\Delta F_1+E\big)\psi_1 = 0.
\end{equation}
Furthermore, $F_1$ has the same behaviour as $F$ at infinity, so we get the
following estimates:
\begin{align}
  \label{eq:bounds_coef}
  \|\partial^{\beta}F_{1}\|_{L^\infty({\mathbb R}^{3N})},\, 
  \| F - F_1 \|_{L^\infty({\mathbb
  R}^{3N})},&\,\|\nabla(F-F_1)\|_{L^\infty({\mathbb R}^{3N})} \nonumber
  \\ 
  &\leq
  C(N,Z,\beta)\,, \,\beta\in\N^{3N}, |\beta|>0.
\end{align}

We will first investigate the necessary regularity properties of the
prefactor $e^{F-F_1}\circ T^*$ in $TU_P$ with respect to $x_P$. Then
we will differentiate the equation (\ref{eq:eq_for_psi1}) with respect
to $x_P$ and analyse the regularity of $\partial_{x_P}^{\gamma}(\psi_1
\circ T^*)$ in $TU_P$. 

In the coordinates defined by $T$ we have
\begin{align}
  \label{eq:transf1}
  x_j &= \frac{x_P}{\sqrt{N_1}} + t_j^* x' &\mbox{ for } j \leq N_1,
  \\
  \label{eq:transf2}
  x_j &= t_j^* x' &\mbox{ for } j > N_1.
\end{align}
So
\begin{align*}
  x_j-x_k =  (t_j^* -t_k^*)x' &\mbox{ for } j,k \leq N_1 \mbox{ or } j,k > N_1.
\end{align*}
In particular these last expressions are independent of $x_P$. 
So when we differentiate $F \circ T^*$ with respect to $x_P$, the only
non-vanishing terms come from derivatives of $|x_j|\circ T^*$ with $j
\leq N_1$ and $|x_j -x_k| \circ T^*$ with $j \leq N_1$, $k > N_1$. The
definition of $U_P$ implies that the function ${\mathbf x} 
\mapsto |x_j|$ is smooth with bounded derivatives on $U_P$ for $j \leq
N_1$, and  
${\mathbf x} \mapsto |x_j-x_k|$ is smooth with bounded derivatives on
$U_P$ for $j \leq N_1$, $k > N_1$. Thus 
\begin{align}
  \label{eq:FandF1}
  \|\nabla \partial_{x_P}^{\gamma} (F-F_1) \circ T^*\|_{L^{\infty}(T U_P)} &+
  \|\partial_{x_P}^{\gamma} (F-F_1) \circ T^*\|_{L^{\infty}(T U_P)} \nonumber\\
  &\leq C(\gamma) \mbox{ for all } \gamma \in {\mathbb N}^3.
\end{align}
Furthermore, we get using (\ref{eq:bounds_coef}) that
\begin{equation}
  \partial_{x_P}^{\gamma} \left(e^{F-F_1} \circ T^*\right) \in
  C^{0,1}(T U_P) \cap L^{\infty}(T U_P) \mbox{ for all } \gamma \in
  {\mathbb N}^3. 
\end{equation}
Hence, due to (\ref{eq:ansatz}) and (\ref{eq:FandF1}), it remains to study
the regularity of $\psi_1\circ T^*$. 

In the rest of the proof we will use the following notation:
We will write $\tilde{F},\tilde{F}_1, \tilde{\psi}_1$ instead of
$F\circ T^*, F_1 \circ T^*, \psi_1\circ T^*$. In particular we have
the following relation 
\begin{align*}
  \psi \circ T^* = e^{\tilde{F}-\tilde{F}_1} \tilde{\psi}_1.
\end{align*}

Notice that since the Laplacian is invariant under orthogonal
transformations we have the following equation for $\psi \circ T^*$: 
$$
  -\Delta (\psi \circ T^*) + (V \circ T^*) (\psi \circ T^*) = E (\psi
  \circ T^*). 
$$
Now, once again using the invariance of the Laplacian and
(\ref{eq:choice_of_F}),  
$$
  \Delta \tilde{F} = V \circ T^*,
$$
so we get the following equation for $\tilde{\psi_1}$ (compare with
(\ref{eq:eq_for_psi1})): 
\begin{align}
  \label{eq:eq_for_psi1_T}
  &L \tilde{\psi}_{1} =0,\nonumber\\
  &L=\Delta+2\nabla(\tilde{F}-\tilde{F}_1)\cdot\nabla
  +\big(|\nabla(\tilde{F}-\tilde{F}_1)|^2-\Delta \tilde{F}_1+E\big).
\end{align}

The analysis of $\tilde{\psi}_1$ will be based on the elliptic
regularity result from Proposition~\ref{prop:GT}. We shall proceed by
induction and for this we will need the following open sets indexed by
$l \in {\mathbb N}$: 
\begin{align*}
  U_l = \Big\{ (x_1,\ldots,x_N) \in {\mathbb R}^{3N}\,\Big|\,& |x_j| > 
  \epsilon(1-2^{-(l+1)}) \mbox{ for } j \in P, \\
  &|x_j-x_k| > \epsilon(1-2^{-(l+1)}) \mbox{ for } 
  j \in P, k \in Q \Big\}.
\end{align*}
It is clear that for $l_1 < l_2$ we have
$$
U_P \subset U_{l_2} \subset U_{l_1} \subset U_0.
$$

We will prove the following statement:

\begin{statement}
For all $\gamma \in {\mathbb N}^3$ we have
\begin{araliste}
  \item
  \label{regularity} 
    $\partial_{x_P}^{\gamma} \tilde{\psi}_1  
    \in C^{1,\alpha}(T U_{|\gamma|})$ for all $\alpha \in (0,1)$. 
  \item 
  \label{H2} 
    $\partial_{x_P}^{\gamma} \tilde{\psi}_1 \in 
    W^{2,2}_{\text{loc}}(T U_{|\gamma|})$.
  \item 
  \label{decay} 
    There exists $c=c(\gamma)>0$ such that
\begin{align*}
    |\nabla \partial_{x_P}^{\gamma} \tilde{\psi}_1
    (x_P,x')|+|\partial_{x_P}^{\gamma} \tilde{\psi}_1 (x_P,x')| \leq c
    e^{-\lambda |(x_P,x')|}  
\end{align*}
for all $(x_P,x')\in T U_{|\gamma|}$.
\end{araliste}
\end{statement}

Let us start by noticing that once this statement is established
Lemma~\ref{lem:smoothness} is proved. 

The proof of the statement proceeds by induction with respect to $|\gamma|$. 

In order to prove the statement for $|\gamma| =0$, let us look at the
equation (\ref{eq:eq_for_psi1_T}). 
We get from Proposition~\ref{prop:GT} that 
\begin{align*}
  \tilde{\psi}_{1}\in C^{1,\alpha}(T U_{0}),
\end{align*}
since the coefficients of the equation (\ref{eq:eq_for_psi1_T}) are
bounded on $T U_{0}$, due to (\ref{eq:bounds_coef}). This proves
(\ref{regularity}) for $|\gamma|=0$. 

We next prove that $\tilde{\psi}_{1} \in
W^{2,2}_{\text{loc}}(T U_0)$. This is accomplished as follows:  
If we use that $\tilde{\psi}_{1}\in C^{1,\alpha}(T U_{0})$ and that
the coefficients in the equation (\ref{eq:eq_for_psi1_T}) (derivatives
of $\tilde{F}-\tilde{F}_1$)  are bounded (again using
(\ref{eq:bounds_coef})), then it is easily seen from 
(\ref{eq:eq_for_psi1_T}) that
$$
  \Delta \tilde{\psi}_1 \in L^2_{\text{loc}}(TU_0).
$$
Therefrom, we get via standard elliptic regularity results (see for
instance Folland~\cite[Theorem 6.33]{Folland}) that $\tilde{\psi}_1 \in
W^{2,2}_{\text{loc}}(T U_0)$. 

Next, we verify the exponential decay estimate (\ref{decay}) for $|\gamma|=0$. 
We know from the assumption (\ref{eq:decay}) that $\tilde{\psi}_1$
decays exponentially, but we also need to prove it for $\nabla
\tilde{\psi}_1$. 
This is done exactly as in the induction step below, using the
exponential decay estimate (\ref{eq:decay}). In order not to repeat
the argument, we refer the reader to the induction step below. 

Suppose now that we have proved (\ref{regularity})-(\ref{decay}) for
all $\gamma$ with $|\gamma| \leq k$. Take a $\gamma$ with length
$|\gamma| = k+1$. 

Differentiating the equation (\ref{eq:eq_for_psi1_T}) for
$\tilde{\psi}_1$ we get the 
following equation for $\partial_{x_{P}}^{\gamma}\tilde{\psi}_1$ (in
the sense of distributions): 
\begin{align}
  \label{eq:paralel_diffentiated}
  &L(\partial_{x_{P}}^{\gamma}\tilde{\psi}_{1})=f_{\gamma},\nonumber\\
  &L=\Delta+2\nabla(\tilde{F}-\tilde{F}_1)\cdot\nabla
  +\big(|\nabla(\tilde{F}-\tilde{F}_1)|^2-\Delta \tilde{F}_1+E\big),\nonumber\\
  &f_{\gamma}=-\sum_{\sigma+\mu = \gamma, |\mu| < |\gamma|} 
  \Big(
    2\left[ \partial_{x_P}^{\sigma}\big(\nabla
  (\tilde{F}-\tilde{F}_1)\big) \right] 
  \cdot \nabla \partial_{x_P}^{\mu} \tilde{\psi}_1 \nonumber\\ 
  &
  \quad\quad
  +\left[ \partial_{x_P}^{\sigma}\big(|\nabla(\tilde{F}-\tilde{F}_1)|^2
           -\Delta \tilde{F}_1+E  \big)
  \right] \partial_{x_P}^{\mu} \tilde{\psi}_1 
  \Big).
\end{align}
In $f_{\gamma}$ only partial derivatives of $\tilde{\psi}_1$ of length
$\leq k$ occur. 

It is clear that in (\ref{eq:FandF1}) we may replace $TU_P$ by
$TU_0$. Therefore we get, 
using the induction hypothesis for (\ref{regularity}), that $f_{\gamma}
\in L^{\infty}(TU_{|\gamma|-1})$. 
A priori $\partial_{x_{P}}^{\gamma}\tilde{\psi}_1$ only satisfies the
equation (\ref{eq:paralel_diffentiated}) in the 
distributional sense. 
In order to apply Proposition~\ref{prop:GT} we need that
$\partial_{x_{P}}^{\gamma}\tilde{\psi}_1 \in
W^{1,2}_{\text{loc}}(U_{|\gamma|-1})$. 
However, this follows from the induction hypotheses for (\ref{H2}), and
the definition of Sobolev spaces.  
Now, due to (\ref{eq:FandF1}), the coefficients of $L$ are bounded,
and therefore we get from Proposition~\ref{prop:GT} that 
$\partial_{x_P}^{\gamma}\tilde{\psi}_1 \in
C^{1,\alpha}(TU_{|\gamma|-1})$ for all $\alpha \in (0,1)$, proving
(\ref{regularity}).
This together with (\ref{eq:FandF1}) implies that $f_{\gamma} \in
L^2_{\text{loc}}(TU_{|\gamma|-1})$. Therefore, we get (invoking once
again that $\partial_{x_P}^{\gamma}\tilde{\psi}_1 \in 
C^{1,\alpha}(TU_{|\gamma|-1})$ for all $\alpha \in (0,1)$ and
(\ref{eq:FandF1})),  
$$
  \Delta (\partial_{x_{P}}^{\gamma} \tilde{\psi}_1) \in
  L^2_{\text{loc}}(TU_{|\gamma|-1}). 
$$
As before, via a standard elliptic regularity theorem
(e.g. Folland~\cite[Lemma 6.32]{Folland}), we see that
$\partial_{x_{P}}^{\gamma} \tilde{\psi}_1 \in
W^{2,2}_{\text{loc}}(TU_{|\gamma|-1})$, and therefore (\ref{H2}) is
proved.  

Finally, we prove the exponential decay of $\partial_{x_{P}}^{\gamma}
\tilde{\psi}_1$. 
Let us write $\epsilon_{l} = \epsilon/2^{l+3}$. Then we
have for all ${\mathbf x} \in U_l$ that
$\overline{B({\mathbf x},2\epsilon_l)} \subset U_{l-1}$.

Since the equation (\ref{eq:paralel_diffentiated}) is
satisfied on $TU_{|\gamma|-1}$ it is in particular satisfied on
$TB({\mathbf x},2\epsilon_{|\gamma|})$ for any ${\mathbf x} \in U_{|\gamma|}$. 
Applying Proposition~\ref{prop:GT} with 
$\Omega' = TB({\mathbf x},\epsilon_{|\gamma|})$, 
$\Omega = TB({\mathbf x}, 2\epsilon_{|\gamma|})$, we get that
\begin{align*}
  | \partial_{x_P}^{\sigma} \tilde{\psi}_1 |_{C^{1,\alpha}(\Omega')}
  \leq C\left( \sup_{\Omega} |\partial_{x_P}^{\sigma}
  \tilde{\psi}_1|+ \sup_{\Omega}| f_{\sigma} | \right).
\end{align*}
Notice, using (\ref{eq:FandF1}), that the coefficients in the
differential operator $L$ are uniformly bounded on all of
$TU_0$. Furthermore, the derivatives of $\tilde{F}$ and $\tilde{F}_1$ in
the expression for $f_{\gamma}$ are also uniformly bounded on 
$U_0$---once again using (\ref{eq:FandF1}). Therefore, using the induction
hypothesis (\ref{decay}), we get for all $(x_P,x')\in TU_{|\gamma|}$: 
\begin{align*}
  |\nabla \partial_{x_P}^{\sigma} \tilde{\psi}_1(x_P,x')| & \leq
  | \partial_{x_P}^{\sigma} \tilde{\psi}_1 |_{C^{1,\alpha}(\Omega')} \\
  &\leq C\left( \sup_{\Omega} |\partial_{x_P}^{\sigma}
  \tilde{\psi}_1|+ \sup_{\Omega}| f_{\sigma} | \right) \\
  & \leq C \sup_{ (y_P,y')\in TB((x_P,x'),2\epsilon_{|\gamma|})}
  e^{-\lambda |(y_P,y')|} \\ 
  & = C e^{-\lambda |(x_P,x')|}.
\end{align*}
In this last line the constant $C$ does not depend on the position
$(x_P,x')$ of the ball, since the coefficients of the equation
(\ref{eq:paralel_diffentiated}) are uniformly bounded on all
$TU_0$. This proves (\ref{decay}) and thus finishes the induction. 
\end{proof}

\begin{remark}
  In the case of molecules we modify $F$ and $F_1$ as follows:
\begin{align*}
  F= \sum_{l=1}^L \sum_{j=1}^N{}-\frac{Z_l}{2} |x_j - R_l| + \sum_{1\leq
  j< k\leq N}  
  \frac{1}{4}|x_j-x_k|,
\end{align*}
and
\begin{align*}
  F_1 = \sum_{l=1}^L \sum_{j=1}^N{}-\frac{Z_l}{2} \sqrt{|x_j-R_l|^2+1} +
  \sum_{1\leq j<  
  k\leq N} \frac{1}{4}\sqrt{|x_j-x_k|^2+1}.
\end{align*}
The rest of the proof is analogous.
\end{remark}

\section{The proof of Theorem~\ref{thm:density_smooth}}

As noted in Remark~\ref{rem:simplify} it is enough to prove smoothness
of each individual term in~\eqref{rho}. We therefore redefine $\rho$
by 
$$
  \rho(x) = \int \psi^2(x,x_2,\ldots,x_N)\,dx_2\cdots dx_N.
$$

Lemma~\ref{lem:smoothness} will be essential in order to prove the
smoothness of $\rho$. 
It suffices to prove that $\rho \in C^{\infty}({\mathbb R}^3 \setminus
\overline{B(0,R)})$ for all $R >0$. 
Therefore, let us assume that $|x|>R>0$.

\begin{remark}
  In the case of molecules we assume
  \begin{align*}
    \min_{1\leq l \leq L} |x - R_l| > R > 0,
  \end{align*}
  and prove that $\rho \in C^{\infty}({\mathbb R}^3 \setminus \left(
  \cup_{l=1}^L\overline{B(R_l,R)}\right))$. 
\end{remark}

Let $\chi_1,\chi_2$ be a partition of unity in ${\mathbb R}_{+}$: 
$\chi_1+\chi_2 =1$, $\chi_1(x) = 1 $ on $[0,R/(4N)]$, $\supp \chi_1 
\subset [0,R/(2N)]$ and
$\chi_j \in C^{\infty}({\mathbb R_{+}})$ for $j=1,2$.

We combine the $\chi_j$'s to make a partition of unity in 
${\mathbb R}^{3N}$. Obviously:
\begin{align*}
  1 = 
  \prod_{1\leq j<k\leq N}(\chi_1+\chi_2)(|x_j-x_k|).
\end{align*}
Multiplying out the above product, we get sums of products of 
$\chi_1$'s and $\chi_2$'s. We introduce the following index sets to 
control these sums:
Define first
\begin{align*}
  M=\{ (j,k) \in \{ 1,\ldots,N\}^2\,|\, j <k \},
\end{align*}
and let
\begin{eqnarray*}
   I &\subset& M, \\
   J & = &M\setminus I.
\end{eqnarray*}
Now define, for each pair $I,J$ as above,
\begin{align*}
  \phi_{I}({\mathbf x}) &= \left(
  \prod_{(j,k)\in I} \chi_1(|x_j-x_k|) \right) 
  \left(\prod_{(j,k)\in J} \chi_2(|x_j-x_k|) \right).
\end{align*}
Then we get
\begin{align*}
  1 = \prod_{1\leq j<k\leq N}(\chi_1+\chi_2)(|x_j-x_k|) =
  \sum_{I\subset M} \phi_{I}({\mathbf x}),
\end{align*}
where the sum is over all subsets $ I \subset M$.

Therefore we obtain:
\begin{align*}
  \rho(x_1) &=
  \int_{{\mathbb R}^{3(N-1)}} \psi^2(x_1,x_2,\ldots,x_N)\,dx_2 \cdots\,dx_N 
  \nonumber \\
  &=
  \sum_{I\subset M} \int \psi^2(x_1,x_2,\ldots,x_N) 
  \phi_{I}(x_1,x_2,\ldots,x_N)\,dx_2 \cdots\,dx_N \nonumber \\
  &\equiv \sum_{I\subset M} \rho_I(x_1).
\end{align*}
We will prove smoothness of each of the individual $\rho_I$'s, so let
us pick and fix an arbitrary $I \subset M$. 
Our strategy is to associate to \(I\) a subset \(P\) of
\(\{1,\ldots,N\}\) such that Lemma~\ref{lem:smoothness} is applicable,
i.e.\ such that $\supp \phi_I \subset U_P$ with $U_P$ defined in
(\ref{eq:def_U_P}) with a suitable $\epsilon > 0$. 
\begin{remark}
To motivate the determination of $P$ and hence the coordinate $x_P$
with respect to which we are allowed to differentiate, 
let us consider 
the following example: Let $N=3$ and
$I=\{(1,2),(2,3)\}$ and $J = \{(1,3)\}$.
Then we have on the support of $\phi_I$ that $|x_1-x_2| \leq
\tfrac{R}{6}$ (due to the $\supp \chi_1$) and since $|x_1| > R,$
$|x_2| \geq \tfrac{5R}{6}$. Further, due to the $\supp \chi_2$, we
have $|x_1-x_3| \geq \tfrac{R}{12}$. Suppose now, we would choose $P =
\{ 1,2\}$ and $Q= \{3\}$, then, according to (\ref{eq:def_U_P}), 
$$
  U_P = \{ (x_1,x_2,x_3) \big| |x_1|>\epsilon, |x_2|>\epsilon,
  |x_1-x_3|>\epsilon, |x_2-x_3|>\epsilon \}, 
$$
for some $\epsilon > 0$.
But then $\supp \phi_I \not \subset U_P$, since $\supp \phi_I$
contains points with $x_2=x_3$. On the other hand one easily checks
that the choice $P=\{1,2,3\}$ is the right one.  
This example shows that we cannot just choose $P$ to be $\{1\} \cup
\{j \leq N \big| (1,j) \in I \}$. On the other hand $P$ cannot be too
big: For $N=3$, $I=\{(1,2)\}$, $J=\{(1,3),(2,3)\}$ it is easily seen
that with $P = \{1,2,3\}$, $\supp \phi_I$ contains points with $x_3=0$
and therefore $\supp \phi_I \not \subset U_P$. 

Physically speaking, we divide the $N$ electrons into $2$
clusters. The electrons $j$ with $j \in P$ define the `maximal
cluster' of electrons containing the electron $1$. This will be done
via an equivalence relation below. Note that the (three-dimensional)
variable $x_{P} =  
\tfrac{1}{\sqrt{|P|}}\sum_{j \in P} x_j$ is (up to a scalar multiple)
the centre (centre  
of mass) of the maximal cluster.  
\end{remark}

Let $\sim$ denote the equivalence relation on $\{1,\ldots,N\}^2$
generated by $I$ and let $P$ denote the equivalence class of $1$.  

Explicitly this means that $j \sim k$ if either $j=k$ or there exists a
sequence $j_1,\ldots,j_l$ with
$j_{s}\in\{1,\ldots,N\}$ for $1\le s\le l$ and with $j_s \neq j_t$ for
$s\neq t$, such that 

\begin{latinliste}
  \item
  \(\quad(j,j_1) \in I\) or \((j_1,j) \in I\), 
  \item
  \(\quad(j_{s},j_{s+1}) \in I\) or  \((j_{s+1},j_{s}) \in I\), for $
  1 \leq s \leq l-1$, 
  \item 
  \(\quad(j_l,k) \in I\) or \((k,j_{l}) \in I\).
\end{latinliste}

Clearly
\begin{equation}
  \label{eq:sequence_length}
  l \leq N-2.
\end{equation}

Thus $P= \{ j \,|\, j \sim 1 \}$, $Q = \{1,\ldots, N\} \setminus P$.
Notice that $P \neq \emptyset$.
In order to be able to apply Lemma~\ref{lem:smoothness}, we have to
show that $\supp \phi_I \subset U_P$ (with a suitable choice of
$\epsilon$ in the definition of $U_P$).  

Let $j \in P$, then $j \sim 1$ and we can choose a sequence
$j_1,\ldots,j_l$ according to the above.  
Taking into account $|x_1| > R$ we have
$$
  |x_j| \geq R - |x_1-x_j|.
$$
Further, with 
$$
  |x_1 - x_j| \leq |x_1-x_{j_1}| + \sum_{s=1}^{l-1}
  |x_{j_s}-x_{j_{s+1}}| + | x_{j_l}-x_j|, 
$$
the length scale of the cut-off's and (\ref{eq:sequence_length}) 
we obtain that
\begin{align}
\label{eq:support1}
  j \in P \Rightarrow \supp \phi_{I}  \subset \{ (x_1,\ldots,x_N) \in 
  {\mathbb R}^{3N} \, \big| \, |x_j| > R/4\}.
\end{align}

\begin{remark}
  In the case of molecules we get
  \begin{align*}
    j \in P \Rightarrow \supp \phi_{I}  \subset \{ (x_1,\ldots,x_N) \in 
    {\mathbb R}^{3N} \, \big| \, \min_{1\leq l \leq L}|x_j - R_l| > R/4\}.
  \end{align*}
\end{remark}

Furthermore, suppose $j \in P$, $k \in Q$, then it is clear that 
$(j,k) \in J$ or $(k,j) \in J$
(because if $(j,k) \in I$, then $1\sim j\sim k$ and therefore $k \in P$),
and therefore:
\begin{align}
  \label{eq:support2}
  j \in P, k\in Q \Rightarrow \supp \phi_{I}  \subset \{ (x_1,\ldots,x_N) 
  \in {\mathbb R}^{3N} \, \big| \,|x_j-x_k| > R/(4N)\}.
\end{align}
Using (\ref{eq:support1}) and (\ref{eq:support2}) we see that $\supp
\phi_I \subset U_P$, with $\epsilon = \tfrac{R}{4N}$ in the definition
(\ref{eq:def_U_P}) of $U_P$. 
Hence we get from Lemma~\ref{lem:smoothness} that $\psi\circ T^*$ is 
(infinitely often) differentiable with respect to the coordinate $x_P 
= \tfrac{1}{\sqrt{|P|}}\sum_{j\in P} x_j$ on the support of
$\phi_{I}\circ T^*$. 

Denote $g_I= \psi^2 \phi_I$, 
and note that all partial derivatives of $\phi_I$ are bounded. We get
from Lemma~\ref{lem:smoothness} that 
\begin{align}
  \label{eq:diff_and_decay}
  &\partial_{x_P}^{\gamma} (g_I \circ T^*) \in C^{0,1}({\mathbb
  R}^{3N}), \nonumber \\ 
  &|\partial_{x_P}^{\gamma} (g_I \circ T^*) (x_P,x')| \leq c_{\gamma}
  e^{-\lambda |(x_P,x')|} \mbox{ for all } (x_P,x') \in {\mathbb
  R}^{3N}. 
\end{align}

Now, we are ready to prove the smoothness of the electron density.
We calculate (using the notation from (\ref{eq:T_stjerne}), and
\eqref{eq:transf1} and \eqref{eq:transf2}): 
\begin{eqnarray}
  \label{eq:delta}
  \rho_I(x)  &=&\int_{{\mathbb R}^{3N-3}}
  g_I(x,x_2,\ldots,x_N)\,dx_2\cdots\,dx_N \nonumber\\ 
  &=&
  \int_{{\mathbb R}^{3N}}
  g_I(x_1,\ldots,x_N)\delta(x-x_1)\,dx_1dx_2\cdots\,dx_N \nonumber\\ 
  &=&
  \int_{{\mathbb R}^{3N}} (g_I\circ T^*)(x_P,x')
  \delta(x-\tfrac{x_P}{\sqrt{N_1}}-t_1^*x') \,dx_P\,dx'\nonumber\\ 
  &=&
  \int_{{\mathbb R}^{3N-3}} (g_I\circ T^*)(\sqrt{N_1}(x-t_1^*x'),x')  \,dx'
\end{eqnarray}

Using (\ref{eq:diff_and_decay}) and Lebesgue integration theory, we
obtain via the chain rule: 
\begin{align*}
  \partial_x^{\gamma} \rho_I(x) = 
  & \partial_{x}^{\gamma} \Big(\int \big(g_I\circ
  T^*\big)\big(\sqrt{N_1}(x-t_1^*x'),x'\big)  \,dx' \Big)\\ 
  =& \big( \sqrt{N_1} \big)^{|\gamma|} \int
  \big(\partial_{x_P}^{\gamma}(g_I\circ T^*)\big) 
  \big(\sqrt{N_1}(x-t_1^*x'),x'\big)  \,dx'. 
\end{align*}
This proves that $\rho$ is smooth away from the nucleus.

The exponential decay of the derivatives of $\rho_I$ is a consequence of
(\ref{eq:diff_and_decay}). This can be seen by a similar calculation as in 
(\ref{eq:delta}) but in reversed order:
\begin{eqnarray*}
  |\partial_x^{\gamma} \rho_I(x)|
  &\leq & \big( \sqrt{N_1} \big)^{|\gamma|} \int \left|
  \big(\partial_{x_P}^{\gamma}(g_I\circ T^*)\big)
  \big(\sqrt{N_1}(x-t_1^*x'),x'\big) 
  \right|\,dx' \\ 
  &\leq & c \int  e^{-\lambda |(\sqrt{N_1}(x-t_1^*x'),x')|} \,dx'\\
  &=&
  c \int (e^{-\lambda |{\mathbf x}|} \circ T^*)(x_P,x')
  \delta\big(x-\tfrac{x_P}{\sqrt{N_1}}-t_1^*x'\big) \,dx_P\,dx'\\ 
  &=&c \int e^{-\lambda |{\mathbf x}|} \delta(x-x_1) \,dx_1\cdots dx_N \\
  &=& c\int e^{-\lambda |(x,x_2,\ldots,x_N)|} \,dx_2\cdots dx_N.
\end{eqnarray*}
Let us write $(x,x_2,\ldots,x_N) = (x,z)$. Then for all $\epsilon \in (0,1)$:
\begin{align*}
  |(x,z)| = (1-\epsilon)|(x,z)|+\epsilon|(x,z)| \geq
  (1-\epsilon)|x|+\epsilon|z|. 
\end{align*}
Therefore
\begin{align*}
  \int e^{-\lambda |(x,z)|}\, dz &\leq 
  e^{-\lambda(1-\epsilon)|x_1|} \int e^{-\lambda \epsilon|z|} \,dz \\ 
  & =  c e^{-\lambda(1-\epsilon)|x_1|}. 
\end{align*}
This verifies inequality (\ref{eq:rho_exponential}) and finishes the
proof of Theorem~\ref{thm:density_smooth}.
\qed

\ %

\begin{acknowledgement}
  The first mentioned author wishes to thank ESI and in particular Prof. 
  Jakob Yngvason for hospitality in the spring 2001.
\end{acknowledgement}

\bibliographystyle{acm}

\end{document}